\documentclass[12pt,a4paper,reqno]{article}
\usepackage{eufrak,bbm,a4}
\newcommand{\frak}{\mathfrak}
\newcommand{\Bbb}{\mathbbm}
\renewcommand{\cal}{\mathcal}
\begin{document}

\begin{center}

{\Large Exact  solutions \\ 
of Reissner-Nordstr\"om type \\
in Einstein-Yang-Mills--systems \\
with arbitrary gauge groups \\
and space-time dimensions}

\vspace{2cm}

{\large Gerd Rudolph and Torsten Tok}

\vspace{1cm}

{ Institute of Theoretical Physics, Leipzig University \\
04109 Leipzig, Augustusplatz 10}

\vspace{1,5cm}

{\large Igor Volobuev}

\vspace{1cm}

{Institute of Nuclear Physics, Moscow State University \\
119899 Moscow, Russia}

\end{center}

\begin{abstract}

We present a class of solutions in Einstein-Yang-Mills-systems 
with arbitrary gauge groups and space-time dimensions, which are
symmetric under the action of the group of spatial rotations. Our
approach is  based on the dimensional reduction method for gauge and
gravitational fields and relates symmetric EYM-solutions to
certain solutions of two-dimensional
Einstein-Yang-Mills-Higgs-dilaton theory. Application of this
method to four-dimensional 
spherically symmetric (pseudo)riemannian space-time yields, in particular,  
new solutions describing both a magnetic and an electric charge in 
the center of a black hole. Moreover, we give an example of a solution
with nonabelian gauge group in six-dimensional space-time. We
also comment on the stability of the obtained solutions. 
 
\end{abstract}

\vspace{2cm}

\section{Introduction}

The last few years were marked by an ever growing interest in solutions of 
Einstein-Yang-Mills-systems  
\cite{Bartnik,GaltsovV,GaltsovE,Bizon,Kunzle,GaltsovV2,GaltsovV3} and in 
the  problem of their stability
\cite{Straumann494}. Most of these solutions were found numerically for the
case of the gauge group $SU(2)$. In fact, the only known analytical
solutions are the Reissner-Nordstr\"om-type solutions, which are
effectively abelian. Such solutions were found for gauge groups $SU(2)$
and $SU(3)$ in \cite{Bartnik,GaltsovV3}. It is believed that such
solutions can be easily constructed for arbitrary gauge groups from
solutions of Einstein-Maxwell-systems by a procedure proposed in
\cite{Yasskin}.

In this paper, we present a class of Reissner-Nordstr\"om-type solutions in
EYM-theories with arbitrary gauge groups and in arbitrary spacetime
dimensions, which are more general than those of ref. \cite{Yasskin}. In
fact, if the spacetime dimension is greater than four, our solutions can
be nonabelian, in contrast to the solutions constructed in accordance
with \cite{Yasskin}. The specific structure of spherically symmetric
four-dimensional spacetime stipulates that our solutions are in this case
abelian. But nevertheless they are more general than those of ref.
\cite{Yasskin}, because their effective gauge group is $ U(1)
\times U(1) $. In particular we find a new solution describing 
both a magnetic and an electric charge in the center of a black hole. 

Our approach is based on the dimensional reduction method for gauge and
gravitational fields and relates symmetric EYM-solutions in spacetime
of arbitrary dimension to certain solutions  of two dimensional
Einstein-Yang-Mills-Higgs-dilaton (EYMHd) theory. This method essentially
simplifies the problem of finding symmetric solutions, and we think that
it can be useful for finding symmetric solutions of more general type.
Moreover, relating the original system to a 2-d EYMHd-system  gives the 
possibility to discuss some aspects of stability of these solutions 
in a quite transparent manner.

\section{EYM-systems with arbitrary gauge group}

We consider Einstein-Yang-Mills-theory with compact gauge group $ G $ on
a spacetime $ E $ of the form
\begin{equation}
E = M \times K/H \quad ,
\end{equation}
where $ M $ is a contractible two dimensional manifold and $
K/H $ is an irreducible compact symmetric space (obviously, the
sphere $ S^n$ is a special case of it). The group $ K $ is a
simple compact Lie group, and its  action on $ E $
is given by the canonical left action of $ K $ on $ K / H $, we denote it by
\begin{equation}
\label{actionE}
\delta : K \times E \rightarrow E \quad .
\end{equation}
Further we assume the existence of a lift
\begin{equation}
\label{actionP}
 \sigma : K \times P \rightarrow P
\end{equation}
of the action
$ \delta $ to automorphisms of the principal bundle $ P ( E , G ) $.

A K-invariant EYM-configuration on $ E $ is given by a metric $ g $ on $E$
and a connection form $ \omega $ on $ P(E,G) $ with
\begin{eqnarray}
{\delta_k}^\ast g & = & g \quad \mbox{and} \\
{\sigma_k}^\ast \omega & = & \omega \quad \forall k \in K \quad .
\end{eqnarray}
First we note that the most general form of  $ K $-invariant metric on
$ E = M \times K/H $ is \cite{Jadczyk}
\begin{equation}
g = h \oplus e^{2 \rho } \gamma_{K/H} \quad ,
\end{equation}
where $ h $ is an arbitrary metric on $ M $, $ \rho $ is a function on $M$ 
and $ \gamma_{K/H} $ is a $ K $-invariant metric on $ K / H $.

If we fix a local coordinate system $ ( x^\mu ) $ in
$ M  $, the metric $ h $ can be written in the form
\begin{equation}
h = h_{\mu \nu} d x^\mu \stackrel{s}{\otimes} d x^\nu  \quad .
\end{equation}

Denoting $ dim K/H = n$, calculating the scalar curvature of $E$
with this metric and omitting a complete divergence, we get

\begin{equation}
\label{E-action}
S_g = \frac{v_{K/H}}{16 \pi  \kappa}  \int_{M}
        \left\{ R_{M} + e^{-2 \rho} R_{K/H} +
        n (n-1) \partial_\mu \rho \partial^\mu \rho \right\}
        \, e^{n \rho} \, \sqrt{\mid h \mid} \, d^2 x
\end{equation}
as the reduced gravitational action, where $v_{K/H}$ is the
volume of the symmetric space $K/H$, calculated with the metric
$\gamma_{K/H}$.

Next we consider a $K$-invariant connection form $ \omega $ on $ P(E,G) $.
Bundles $ P( E,G ) $, admitting lifts ( \ref{actionP} ) of the action
( \ref{actionE} ), are in one-to-one correspondence with pairs $ (\lambda,
\hat P) $ \cite{KMRV,Rud1}, where
\begin{equation}
\lambda : H \rightarrow G
\end{equation}
is a group homomorphism and $ \hat P $ is a principal bundle over $ M $
with reduced gauge group $ C = C_G ( \lambda (H)) $
(centralizer of $\lambda ( H ) $ in $G$ ). A $K$-invariant connection form
$\omega$ on $ P ( E,G) $ is in one-to-one correspondence  
\cite{KMRV,FM,HST,Rud2} with a
pair $ (\hat \omega , \Phi ) $, where $ \hat \omega $ is a
connection form on $ \hat P $ and $ \Phi $ is a mapping
\begin{equation}
\Phi : \hat P \rightarrow {\frak M}^\ast \otimes \frak G \quad ,
\end{equation}
satisfying
\begin{equation}
\label{phi-map}
Ad{\lambda ( h )} \circ \Phi ( \hat p ) = \Phi ( \hat p ) \circ
Adh \quad , \, \forall h \in H \quad .
\end{equation}
Here $ \frak G $ is the Lie algebra of $G$, and $ \frak M $ is the
complement of the Lie algebra $ \frak H $ in the reductive
decomposition $ \frak K =  \frak H \oplus  \frak M$.

To express the Yang-Mills action in terms of fields on $ M $,
we take a section $ s $ in the bundle $ \hat P $ and set
$ \hat A := s^ \ast \hat \omega $ , $ \varphi  := s^\ast \Phi $.
Let $ \{e_i\} $ be a basis of $ \frak M $, orthonormal with respect
to $ \gamma_{K/H} $. The pure Yang-Mills action on $ E $ reduces due to
$K$-invariance to the following action on $ M $
\begin{eqnarray}
S_{YM} = \frac{v_{K/H}}{g^2} \int_{M} \bigg( <\hat F_{\mu \nu},
\hat F^{\mu \nu} > +
2 e^{- 2 \rho} \sum_i <D_\mu \varphi(e_i) , D^\mu \varphi(e_i)> - \nonumber \\
\label{YM-action}
- e^{- 4 \rho} V ( \varphi ) \bigg) e^{n \rho} \,
\sqrt{\mid h \mid } d^2 x \quad .
\end{eqnarray}
Here $ < , > $ is the canonically normalized $ Ad$-invariant
bilinear form on $ \frak G $, $ \hat F $ is 
the curvature of $ \hat A $, $ D_\mu
\varphi(e_i) = \partial_\mu\varphi(e_i) + [\hat A_\mu ,
\varphi(e_i)] $ is the
covariant derivative of $ \varphi $ with respect to $ \hat A $,
and the scalar field potential is given by
\begin{eqnarray}
\label{orbitcurv}
V( \varphi ) & = & - \sum_{k , l} < F_{k l} , F_{k l} > \quad , \\
\label{orbitcurv1}
F_{k l} & = & [ \varphi(e_k) , \varphi (e_l) ] - \varphi ([e_k,e_l] \downarrow
\frak{ M }) - \lambda' (  [e_k,e_l] \downarrow \frak{ H} ) .
\end{eqnarray}
Obviously the second term of ( \ref{orbitcurv1} ) vanishes for symmetric
spaces $ K/H $, and one can prove that $ V(\varphi) $ in this case
is a Higgs potential \cite{KMRV}.

Thus, the reduced action of the full EYM-system is:
\begin{eqnarray}
\label{EYM-action}
S &=& \int_{M} \big[ \frac{1}{g^2} ( <\hat F_{\mu \nu},
\hat F^{\mu \nu} > + 2 e^{- 2 \rho}  <D_\mu \varphi , D^\mu \varphi>
- e^{- 4 \rho} V ( \varphi ) ) \nonumber \\
&& + \frac{1}{16 \pi  \kappa}
        \left( R_{M} + e^{-2 \rho} R_{K/H} +
        n (n-1) \partial_\mu \rho \partial^\mu \rho \right) \big]
        \, e^{n \rho} \, \sqrt{| h |} \, d^2 x
\end{eqnarray}
Clearly, this is the action of 2-dimensional dilaton gravity
coupled to gauge and Higgs fields.

We are interested in $K$-symmetric solutions of EYM-theory. One
can prove that there is a one-to-one correspondence between
$K$-symmetric solutions of EYM-theory and solutions of the field equations
resulting from the reduced action \cite{Jadczyk,Coleman}.
Therefore, it is sufficient to solve the field equations of the
reduced theory. These equation are obtained by the
variation of the reduced  action with respect
to the field variables.
 Note that we have to vary the action
(\ref{EYM-action}) before choosing particular coordinates, otherwise
we lose some field equations.

We set $x^1 = r = \exp{ \rho} $ and choose $ x^0 = t $ in such a
way, that the metric tensor
$ (h_{\mu \nu})$ is diagonal. Let $ \alpha^2 = \varepsilon h_{0
0} $ and $\beta^2 = h_{1 1}$,
where the variable $\varepsilon$ has value 1 in Euclidean and value
$-1$ in Minkowski spacetime. Then the resulting equations are
\begin{eqnarray}
\label{var-V}
0 = g^2 \frac{\delta S}{\delta \varphi(e_l)}  &=&
       - 4 r^{(n-4)} \alpha \beta
        \sum_k [ \varphi(e_k) , [ \varphi (e_k) , \varphi(e_l) ]
        - \lambda' ( [e_k , e_l] ) ]
         \nonumber \\
\label{var-phi}
&&    - 4 D_\mu \left( \alpha \beta r^{(n-2)}
      D^\mu \varphi(e_l) \right)  \\
\label{var-A}
0 = g^2 \frac{\delta S}{\delta \hat A_\nu} &=&
      - 4 D_\mu \left( \alpha \beta r^{n} \hat F^{\mu \nu} \right) +
      4 \alpha \beta r^{n} \sum_k [ \varphi (e_k) , D^\nu \varphi(e_k) ]
     \\
\label{var-h01}
0 = \frac{1}{\alpha \beta} \frac{\delta S}{\delta h^{0 1}} &=&
     =   \frac{1}{16 \pi  \kappa} \frac{\partial_0 \beta}{\beta}
        n r^{(n-1)}
      + \frac{2}{g^2} r^{(n-2)}  <D_0 \varphi , D_1 \varphi>
       \\
\nonumber
0 =  \frac{1}{\alpha \beta} \frac{\delta S}{\delta e^{\rho}} & =&
       \frac{1}{16 \pi  \kappa} \bigg[ n r^{(n-1)} R_{M}
      + (n-2) r^{(n-3)} R_{K/H} \\
\nonumber
&&      - n (n-1) (n-2) r^{(n-3)} \frac{1}{\beta^2}
      - 2 n (n-1) r^{(n-2)} \partial_1 \left(\frac{\alpha}{\beta}\right)
      \frac{1}{\alpha \beta} \bigg] + \\
\nonumber
&&     \frac{1}{g^2} \bigg[ n r^{(n-1)} < \hat F_{\mu \nu} , \hat F^{\mu \nu} > +
       2 (n-2) r^{(n-3)}  <D_\mu \varphi,D^\mu \varphi > \\
\label{var-r}
&&       - (n-4) r^{(n-5)} V(\varphi) \bigg] \\
\nonumber
0 = \frac{\delta S}{\delta \alpha} &=&
      \frac{1}{16 \pi  \kappa} \bigg[
       + 2 n r^{(n-1)} \frac{\partial_1 \beta}{\beta^2}
       + r^{(n-2)} \beta R_{K / H}
       - n (n-1) r^{(n-2)} \frac{1}{\beta} \bigg] \\
\nonumber
&&    + \frac{1}{g^2} \bigg[
       - 2 \frac{\varepsilon}{\alpha^2 \beta} r^n < \hat F_{0 1} , \hat F_{0 1} >
       -\frac{2 \varepsilon \beta}{ \alpha^2} r^{(n-2)} <D_0 \varphi , D_0
         \varphi > \\
\label{var-alpha}
&&       + \frac{2 }{ \beta} r^{(n-2)} <D_1 \varphi , D_1 \varphi >
       - \beta r^{(n-4)} V ( \varphi ) \bigg] \\
\nonumber
0 = \frac{\delta S}{\delta \beta} &=&
      \frac{1}{16 \pi  \kappa} \bigg[
       - 2 n r^{(n-1)} \frac{\partial_1 \alpha}{\beta^2}
       + r^{(n-2)} \alpha R_{K / H}
       - n (n-1) r^{(n-2)} \frac{\alpha}{\beta^2} \bigg] \\
\nonumber
&&    + \frac{1}{g^2} \bigg[
       - 2 \frac{\varepsilon}{\alpha \beta^2} r^n <  \hat F_{0 1} ,
         \hat F_{0 1} >
       + \frac{2 \varepsilon }{ \alpha} r^{(n-2)} <D_0 \varphi , D_0
         \varphi > \\
\label{var-beta}
&&       - \frac{2 \alpha}{ \beta^2} r^{(n-2)} <D_1 \varphi , D_1 \varphi >
       - \alpha r^{(n-4)} V ( \varphi ) \bigg]
\end{eqnarray}

These equations are difficult to solve in the general case. But they simplify
considerably, if we restrict ourselves to solutions corresponding
to constant {\it and} covariantly constant Higgs fields. In this case 
we have
\begin{equation}
\label{symm-braek}
d \varphi = 0 \quad \mbox{and} \quad [A_i , \varphi] = 0 , 
\quad i = 0 , 1 \quad .
\end{equation}
Inserting (\ref{symm-braek}) into equation (\ref{var-V}) gives 
\begin{equation}
\frac{\partial V}{\partial \varphi(e_l)} 
 \equiv  \sum_k [ \varphi(e_k) , [ \varphi (e_k) , \varphi(e_l) ]
      - \lambda' ( [e_k , e_l] ) ]  = 0 \quad \forall l   \quad .
\end{equation}
Thus, the Higgs field configurations of such
solutions are the extrema of the potential $ V(\varphi)$. 
Equation (\ref{symm-braek}) is, of course, the equation of spontaneous
symmetry breaking for $\varphi $ being in the Higgs vacuum.

Now we solve the obtained system of equations for this special case. 
Equations (\ref{var-alpha} )  and (\ref{var-beta})
give that  $ \alpha \beta $ is independent of $r$, and we can set
\begin{equation}
\label{alpha*beta}
\alpha \beta = \eta (t ) \quad .
\end{equation}
Our next step is to solve the Yang-Mills-equations (\ref{var-A}) in the
special gauge $  \hat A_1 = 0 $. In this gauge we have 
$$ \hat F_{1 0} = \partial_1  \hat A_0 \nonumber \quad $$ 
and we get for the zero component of ( \ref{var-A} )
\begin{equation}
  0 = \partial_1 \left( \frac{r^n}{\alpha \beta}  \hat F_{ 1 0} \right)
       = \partial_1 \left( \frac{r^n}{\alpha \beta} \partial_1  \hat A_0 \right)
\label{YM1} \quad .
\end{equation}
Taking into account equation (\ref{alpha*beta}) we get after
integration of (\ref{YM1}):
\begin{eqnarray}
\partial_1  \hat A_0 &=&  \hat F_{ 1 0 } = \frac{{\cal E}(t) 
\eta(t)}{r^n} \quad ,
\label{F-01} \\
 \hat A_0 &=& - \frac{{\cal E}(t) \eta(t)}{(n-1) r^{(n-1)}} + {\cal F} 
( t ) \quad ,
\end{eqnarray}
where $ {\cal F}(t)$ and $ {\cal E}(t)$ are time dependent elements of the
Lie algebra ${\frak R}$ of the group $R$ of the unbroken symmetry (which
coincides with ${\frak C}$, the centralizer of $\lambda'({\frak H})$ in
${\frak G}$, if $\varphi = 0$). The remaining gauge freedom always allows
us to put $ {\cal F}(t) = 0 $. Then the first component of the
Yang-Mills-equation (\ref{var-A}) leads to
\begin{eqnarray}
0 &=& \partial_0  \left( \frac{r^n}{\alpha \beta}  \hat F_{ 0 1 } \right) +
       [  \hat  A_0 ,  \hat F_{0 1} ]  \frac{r^n}{\alpha \beta} \nonumber \\
&=&  \partial_0  \left( {\cal E} ( t )  \right)
         + \left[ - \frac{{\cal E}(t) \eta(t)}{(n-1) r^{(n-1)}}
         , \frac{{\cal E}(t) \eta(t)}{r^n} \right]  \frac{r^n}{\alpha \beta}
\nonumber \\
&=& \partial_0 ( {\cal E}(t) ) \quad .
\end{eqnarray}
This means that $ {\cal E} $ has to be a constant element of the
Lie algebra $ {\frak R}$. Thus, the gauge field $ \hat A $ has the simple
form
\begin{equation}
\label{Eichfeld-A}
 \hat A =  - \frac{\eta(t) \quad {\cal E}}{(n-1) r^{(n-1)}} dt \quad .
\end{equation}
Another consequence is, that the { \it gauge-invariant} quantity
\begin{equation}
\label{F-quadrat}
C :=
\frac{r^{2 n}}{ \alpha^2 \beta^2} < \hat F_{0 1}, \hat F_{0 1}>
= < {\cal E} , {\cal E} >
\end{equation}
has to be constant.
Taking into account equations (\ref{F-quadrat}) and (\ref{alpha*beta})
we can solve equation (\ref{var-alpha}), which gives
\begin{equation}
\label{beta}
\frac{n r^{(n-1)}}{\beta^2} = \frac{r^{(n-1)}}{n-1} R_{K/H} +
\frac{16 \pi  \kappa}{g^2} \left(
\frac{2 \varepsilon C}{(n-1) r^{(n-1)}} - \frac{V(\tilde \varphi)\, 
r^{(n-3)}}{n-3}
\right) + D \quad ,
\end{equation}
where $V(\tilde \varphi) $ is the value of the potential
(\ref{orbitcurv}) in an extremum $\tilde \varphi $, and $ D $ is
an integration 
constant. Because of equation (\ref{var-h01}),
the quantity $\beta$ has to be independent of t. Hence $ D $ has to be
independent of t, too.

Thus, we arrive at the following result:
for every solution of the EYM-system, corresponding to a
constant and covariantly constant field $\tilde \varphi$, 
the metric tensor is given by
\begin{eqnarray}
\beta^2 &=&  \bigg( \frac{R_{K/H}}{n(n-1)} +
\frac{16 \pi  \kappa}{g^2} \left[
\frac{2 \varepsilon C}{n (n-1)} r^{2(1-n)}
- \frac{V(\tilde \varphi) }{n (n-3) r^2}
\right] + \frac{D}{n r^{(n-1)}} \bigg)^{-1} , \\
\alpha^2 &=& \frac{\eta(t)^2}{\beta^2} \quad .
\end{eqnarray}
Differentiating equation (\ref{var-alpha}) and using equation
(\ref{alpha*beta}) one can easily see that our solutions also
fulfill equation
(\ref{var-r}).
Thus, we really have got a solution of the reduced EYM-system
with the action (\ref{EYM-action}). This solution can be of
interest for  2-dimensional dilaton gravity.  In fact, it is 
similar to the solution in 2-dimensional dilaton gravity coupled
to abelian gauge field, which was found in \cite{Kiem}.

Since we are interested in the solutions of the original theory, 
now we have to reconstruct the Yang-Mills gauge
field $ A $ on  spacetime $ M \times K / H $.
For this purpose, we define a  section $ s_2 $ in the  principal
$ H $-bundle $ K  \rightarrow K / H  $. Let $ \theta $ be the canonical
left-invariant Lie-algebra-valued 1-form
on the group $ K $, $ \bar \theta : = s_2^\ast \theta $ its pull back, and
$  \bar \theta_{ {\frak H }} $ , $
                 \bar \theta _{ {\frak M }}  $ the 
$ \frak H - $ and $ \frak M $-components  of the latter with
respect to the reductive decomposition $ \frak K = \frak H \oplus \frak M $.
Then the gauge field $ A $ on $ M \times K / H $ is given by \cite{KMRV}
\begin{equation}
\label{gauge field}
A = \hat A + \varphi ( \bar \theta _{ {\frak M }} )  +
          \lambda' ( \bar \theta _{ {\frak H }} ) \quad .
\end{equation}

\section{Examples}

First we consider the four dimensional spherically symmetric space $E
= M \times S^2$ with coordinates $ ( t, r , \vartheta , \phi ) $.
Since $S^2 = SU(2)/U(1)$, the space-time symmetry group $ K $ is
$ SU ( 2 ) $ and the stabilizer group
$ H $ is $ U ( 1 ) $. We parameterize $ SU ( 2 ) $ by the Euler angles
\begin{eqnarray}
SU ( 2 ) \ni u &=& \frac{\sqrt{2}}{2} ( 1 - i \sigma_2 )
\exp{( i \frac{\phi}{2} \sigma_3 )}
\exp{( i \frac{\vartheta}{2} \sigma_2 )}
\exp{( i \frac{\psi}{2} \sigma_3 )} \quad , \\
&& 0 \leq \phi < 2 \pi ,
0 \leq \vartheta \leq \pi , 0 \leq \psi \leq 4 \pi ,
\end{eqnarray}
where $ \sigma_i $ are the Pauli matrices.
The group $ H = U ( 1 ) $ is given by
\begin{equation}
H = \left\{
   \exp{ ( i \frac{\psi}{2} \sigma_3 )} , 0 \leq \psi \leq 4 \pi \right\} .
\end{equation}
Hence, we have the reductive decomposition
$ \frak K = \frak H \oplus \frak M $, with
\begin{eqnarray}
\frak H &=&  \{ z i \sigma_3 ; z \in \Bbb R \} , \\
\frak M &=&  \{ x i \sigma_1 + y i \sigma_2 ; x,y \in \Bbb R \} .
\end{eqnarray}
If we define the section $ s_2 : S^2 = SU ( 2 ) / U ( 1 ) \rightarrow
SU ( 2 ) $ by
\begin{equation}
s_2 ( \vartheta , \phi ) : = \frac{\sqrt{2}}{2} ( 1 - i \sigma_2 )
\exp{( i \frac{\phi}{2} \sigma_3 )}
\exp{( i \frac{\vartheta}{2} \sigma_2 )} ,
\end{equation}
we get
\begin{equation}
s_2^\ast \theta = \frac{i}{2} \sigma_3 \cos{ \vartheta}\, d \phi 
+ \frac{i}{2} \sigma_2 d \vartheta 
+ \frac{i}{2} \sigma_1 \sin{\vartheta}\,  d \phi  .
\end{equation}

The invariant metric $ g $ on $E = M \times S^2$ is given by
\begin{equation}
g = \varepsilon \alpha^2 d t^2 + \beta^2 d r^2 +
r^2 ( d \vartheta^2 + \sin^2{\vartheta}\, d \phi^2 ) .
\end{equation}
We take the metric in the brackets as the standard metric
$\gamma_{K/H}$ on $S^2$.
The scalar curvature $ R_{K/H} $
of the symmetric space $ S^2 = K / H $, calculated with this
metric, is 2; the dimension $ n $ of $ K / H $ is obviously 2.

The action of $ SU ( 2 ) $ on
$ S^2 = SU ( 2 ) / U ( 1 ) $ generates the following mapping $\Sigma $ from
$ \frak M $ onto the tangent space $ T_{(\vartheta = \pi/2 ,
\phi = 0 )} S^2 = T_{[e]} SU(2)/U(1) $ :
\begin{eqnarray}
\Sigma(\frac{1}{2} i \sigma_1) &= &  \partial_\phi  \, , \\
\Sigma(\frac{1}{2} i \sigma_2) &= &  \partial_\vartheta \, ,
\end{eqnarray}
and we see, that $ ( e_1 ,e_2 ) = ( i \sigma_1 /2 ,i \sigma_2 / 2 ) $
is a  basis of $ \frak M $, orthonormal with respect to $ \gamma_{K/H} $.

Now we have to specify the gauge group $ G $ and the homomorphism
$ \lambda :  H \rightarrow  G $ respectively.  We denote  $ \lambda' : 
{\frak H} \rightarrow {\frak G} $ the tangent mapping induced by $ \lambda $.  
We choose $ G = SU ( 4 ) $ as a simple example for demonstrating our method. 
Let $ ( \alpha_1 , \alpha_2 , \alpha_3 ) $ be
a system of simple roots of $ \frak{ G} = su ( 4 ) $. We map
$ i \sigma_3 \in \frak H $ to $ i h_{\alpha_1} + i h_{\alpha_3}$ by
$ \lambda' $, where $ h_{\alpha_k} $ is 
the element of the Cartan
subalgebra of $ su ( 4 ) $ dual to the root $ \alpha_k $ with
respect to  the  canonically  normalized bilinear form on $\frak G $.
One can easily see that  the centralizer
$ {\frak C}_{\lambda'(\frak H )} ({\frak G} ) $ of $ \lambda' ( \frak H ) $
in $ \frak G $ consists of  
two $ { su ( 2 )} $-subalgebras generated by the
 roots $ \alpha_1 + \alpha_2 $ and $ \alpha_2 + \alpha_3 $ and an abelian
one-dimensional subalgebra spanned by the element $ 1 / \sqrt{2}
 ( h_{\alpha_1} + h_{\alpha_3} ) $
of the Cartan
subalgebra. Thus, the gauge group
of the  reduced theory is $ C = SU(2) \times SU(2)  \times U(1) $.

To define the field $ \varphi $, we have to solve equation 
(\ref{phi-map}) for the intertwining operator $ \Phi $. 
There are 4 complex parameters describing the constant Higgs
field $ \varphi $. We set  
\begin{eqnarray}
\varphi ( \frac{1}{2} \sigma_1 + i \frac{1}{2} \sigma_2 ) &=&
a e_{\alpha_1} + b e_{\alpha_3} + c e_{-\alpha_2} + 
d e_{\alpha_1+\alpha_2+\alpha_3} \quad , \\
\varphi ( \frac{1}{2} \sigma_1 - i \frac{1}{2} \sigma_2 ) &=&
\bar a e_{-\alpha_1} + \bar b e_{-\alpha_3} + \bar c e_{\alpha_2} + 
\bar d e_{-\alpha_1-\alpha_2-\alpha_3} \quad ,
\end{eqnarray}
where $ a,b,c,d \in {\Bbb C} $ and bar denotes complex conjugation.
In accordance with equations (\ref{orbitcurv}) and (\ref{orbitcurv1}), 
the potential $ V $ as a function of $ a,b,c,d $ takes the form
\begin{eqnarray}
V &=& - 2 <  F_{\vartheta \phi} , F_{\vartheta \phi} > \\
&=& \frac{1}{2} \big( (a \bar a + c \bar c -1 )^2 + 
   (a \bar a + d \bar d -1 )^2 + ( b \bar b + c \bar c -1)^2 + 
    \\
\nonumber
&& (b \bar b + d \bar d -1 )^2
+  2 ( a \bar c + \bar b d ) (\bar a c + b \bar d ) +
   2 ( a \bar d + \bar b c ) (\bar a d + b \bar c ) \big) .
\end{eqnarray}
Now we have to look for the extrema of the potential $ V $. 
We get 3 typical solutions: first, the trivial solution $ a=b=c=d=0 $;
 second, the solution with $ a=c=d=0 , b=1 $; and 
third, the solution with $ a=b=1 , c=d=0 $.

In the first case $ a=b=c=d=0 $, the field $ \varphi $ vanishes 
and the group $ R $ of the unbroken symmetry is the centralizer 
of $ \lambda( H) $ in $ G $, i.e. $ SU(2) \times SU(2) \times U(1)$. 
The gauge field $ A $ on 
$ E $ is given by 
\begin{eqnarray}
\nonumber
A &=& \hat A + \varphi ( \bar \theta_{\frak M} ) + 
\lambda' (\bar \theta_{\frak H}) \\
&=& - \frac{\eta (t)}{r} {\cal E} dt + \frac{i}{2} ( h_{\alpha_1} + 
h_{\alpha_3} ) \cos{\vartheta}\, d \phi \quad , \quad {\cal E} \in {\frak R} .
\end{eqnarray}
Obviously $ A $ is $ u (1) \oplus u(1) $ valued. The potential $ V $ 
has the value $ 2 $.

In the second case $ a=c=d=0 , b= 1 $, the group of the unbroken symmetry 
is $ U(1) \times U(1) $. The Lie algebra $ \frak R $ is spanned by the 
elements $ h_{\alpha_1} $ and $ 2 h_{\alpha_2} + h_{\alpha_3} $ of 
the Cartan subalgebra of $ \frak G $. The gauge field $A$ on $ E $ is 
given by
\begin{eqnarray}
\nonumber
A &=& \hat A + \varphi ( \bar \theta_{\frak M} ) + 
\lambda' (\bar \theta_{\frak H}) \\ 
&=& - \frac{\eta (t)}{r} {\cal E} dt + 
\frac{i}{2} ( h_{\alpha_1} ) \cos{\vartheta}\, d \phi + 
\big\{\frac{1}{2} (e_{\alpha_3}-e_{-\alpha_3}) d \vartheta + \\
\nonumber
&& \frac{i}{2} (e_{\alpha_3}+e_{-\alpha_3}) \sin{\vartheta}\, d \phi +
\frac{i}{2} ( h_{\alpha_3} ) \cos{\vartheta}\, d \phi \big\} 
 \quad , \quad {\cal E} \in {\frak R} \quad .
\end{eqnarray}
First observe  that $ h_{\alpha_1} $ and $ {\cal E} $ commute with the $ A_1 $ 
subalgebra 
spanned by $ e_{\alpha_3} , e_{-\alpha_3} $ and $ h_{\alpha_3} $
and, second, that the part in the braced brackets is a pure gauge. 
Therefore, we can choose a gauge such that the part in the braced brackets
vanishes and the rest remains unchanged. 
Then we see that the gauge field is effectively 
$ u(1) \oplus  u(1) $ valued. 
The potential $ V $ has the value $ 1 $. 

We can also calculate the first Chern number of the $ U(1)$ connections 
on $ S^2 $, giving the magnetic field component.  
For this purpose we have to consider the injective homomorphism 
$\tau : U(1) \rightarrow SU(4) $ defined by
\begin{equation}
\tau ( e^{i \psi} ) := e^{i \psi h} \in SU(4) \quad ,
\end{equation}
where $ h = h_{\alpha_1} + h_{\alpha_2} $ in the first and
$ h = h_{\alpha_1} $ in the second case. Then we get
\begin{equation}
c = \frac{1}{2 \pi i} \int_{S^2} (\tau')^{-1} ( F ) =
\frac{1}{2 \pi i} \int_{S^2} - \frac{i}{2}  \sin{\vartheta}\, 
d \vartheta \wedge d \phi = - 1 \quad .
\end{equation}
Thus, we can interpret the $ S^2 $ part of the connection as a magnetic 
monopole with strength 1.

The third solution 
corresponds to $ V = 0 $, i.e. to the Higgs vacuum. In this case, 
the homomorphism $ \lambda' :
{\frak H} \rightarrow {\frak G} $  can be extended to a homomorphism 
from $ {\frak K} $ to $ \frak G $. Hence,  
the field strength $ F_{\vartheta \phi} $ 
vanishes, and we have only a $ U(1) $ valued electric field 
\begin{equation}
 F = \frac{\eta(t)}{r^2} {\cal E} d r \wedge d t \quad ,
\end{equation}  
with $ {\cal E} \in {\frak R} $. The group $ R $ of the unbroken symmetry 
in this case is $ SU(2) $, and its Lie algebra $ \frak R $ is spanned 
by $ e_{\alpha_1+\alpha_2} + e_{\alpha_2+\alpha_3}, 
e_{-\alpha_1-\alpha_2} + e_{-\alpha_2-\alpha_3}  $ and 
$ h_{\alpha_1}+ 2 h_{\alpha_2} + h_{\alpha_3} $. 
The fact that the solution corresponding to the Higgs vacuum of
the reduced theory is a pure gauge on $ S^2 $ is valid for an arbitrary
gauge group $ G $. It is due to the specific structure of the
spherically symmetric spacetime $E = M \times K/H $ in four  
dimensions, which includes the two-sphere $ S^2 = SU ( 2 ) / U (
1 ) $, and a general theorem \cite{Dynkin}, which says that any
embedding of the Cartan element $h_{\alpha}$ of $A_1$ into an arbitrary Lie
algebra can be extended to an embedding of the whole $A_1$.

Now we can write down the metric coefficients $ \alpha $ and $ \beta $.
If we set the integration constant $ D = - 4 M $,
they take the form
\begin{eqnarray}
\alpha^2 &=& \eta(t)^2 \bigg\{ 1 + \frac{16 \pi  \kappa}{g^2}
\frac{1}{r^2}
\left( \frac{2 \varepsilon < {\cal E} , {\cal E} >}{2} +
\frac{V({\tilde \varphi}) }{2} \right)
- \frac{2 M}{ r } \bigg\} \quad ,\\
\beta^2 &=&  \left\{ 1 + \frac{16 \pi  \kappa}{g^2}
\frac{1}{r^2}
\left( \frac{2 \varepsilon < {\cal E} , {\cal E} >}{2} +
\frac{V(\tilde \varphi) }{2} \right)
- \frac{2 M}{ r } \right\}^{- 1}  \quad .
\end{eqnarray}

We can interpret this solution in the first two cases as that of 
a magnetic and an electric charge in
the center of  a black hole with Schwarzschild radius $ 2 M $, 
but in the third case we have only an electric charge in the center 
of the black hole. These solutions
are similar to the well known Reissner-Nordstr\"om solution of
Einstein-Maxwell-theory. We stress that 
our solutions can be easily found explicitly for  arbitrary 
gauge groups $G $. One can show - using equation (\ref{symm-braek}) - 
that our approach,  independently of 
the gauge group $G$ and the homomorphism $ \lambda $, 
always gives commutative solutions, if the 
stabilizer group $ H $ is $ U ( 1 ) $. 
Nevertheless, our
solutions are more general than those of \cite{Yasskin}, because
the field $A$  in the first and second cases takes values in a 
two dimensional abelian subalgebra and not in a
one-dimensional one, as in \cite{Yasskin}.

Next we give  an example that our approach can also lead to noncommutative
solutions. We consider the case of six-dimensional space-time $ 
{\Bbb R}^2 \times K / H $, with
\begin{equation}
 K / H = Sp(2) / (Sp(1) \times Sp(1) ) \equiv SO(5) / SO(4) = S^4 .
\end{equation}
Then we have $ K = Sp(2) $ and $ H = Sp(1) \times Sp(1) = SU(2)
\times SU(2) $. 
The embedding of the Lie algebra $su(2) \oplus
su(2)$ into the Lie algebra $sp(2)$ is given by the decomposition
of its defining representation $\underline{4}$:
$$ \underline{4} \longrightarrow (\underline{2},\underline{1}) +
(\underline{1}, \underline{2}) \, , $$
i.e. if $\{e_{\alpha} , e_{\mu} \} , \alpha=1,2 \, , \mu=3,4 $ is 
a basis of $ R_{\underline{4}} $, the representation space of $\underline{4}$, 
then the first $sp(1)$ acts on the space spanned by $ e_1 , e_2 $ in 
the fundamental representation and trivially on the space spanned by 
$ e_3 , e_4 $, the second $sp(1)$ acts correspondingly on $ (e_3, e_4) $ 
in the fundamental representation 
and trivially on $ (e_1,e_2) $. 
It is known that the adjoint representations of $ sp(n)$ and $ sl(n) $ 
can be expressed in 
terms of the their  
fundamental representations by \cite{Dynkin}
\begin{eqnarray}
ad \, sp(n) &=& \underline{(2 n)} \stackrel{s}{\otimes}  \underline{(2 n)} 
\, , \\ 
ad \, sl(n) &=& \underline{n}^* \tilde \otimes \, \underline{n} \, ,
\end{eqnarray}
where tilde means dropping out a one-dimensional trivial representation, 
$ \underline{n^*} $ denotes the contragradient representation, $ t(x)^* = 
- t(x)^T $, and $ \stackrel{s}{\otimes} $ denotes the symmetrized 
tensor product.
Employing the method developed in \cite{RV}, we can easily construct a
basis of generators of $sp(2)$, adapted to the reduction to the $su(2) \oplus
su(2)$ - subalgebra. 
This basis can be chosen to be
$\{h^{\alpha}_{\beta},\, h^{\mu}_{\nu}, \, d_{\alpha \mu}\}, \quad
{\alpha},{\beta}=1,2 , \, {\mu},{\nu} = 3,4$, with
$\{h^{\alpha}_{\beta}=\varepsilon^{\alpha \gamma} e_\gamma 
\stackrel{s}{\otimes} e_\beta \}
,\, h^{\alpha}_{\alpha} = 0; \, \{
h^{\mu}_{\nu}= \varepsilon^{\mu \rho} e_\rho 
\stackrel{s}{\otimes} e_\nu \}, 
\, h^{\mu}_{\mu} = 0$ being the generators of $su(2) \oplus
su(2)$ and $ \{d_{\alpha \mu}=e_\alpha \stackrel{s}{\otimes} e_\mu\}  $ 
being a basis of $ \frak M $. 
The antisymmetric tensor $\varepsilon_{i j} = \varepsilon^{i j},\,
i,j = 1,\ldots ,4 $ is defined by 
$$ \varepsilon_{12} = \varepsilon_{34} =1 , \quad \varepsilon_{13} 
= \varepsilon_{14} = \varepsilon_{23}= \varepsilon_{24} = 0. $$
We use the 
convention that the indices are raised by the second index and
lowered by the first one, i.e. $\varepsilon^{\alpha \beta}
d_{\beta \mu} = {d^\beta}_\mu $. 

The generators of $sp(2)$ have the following commutation relations
\begin{eqnarray}\label{com}
\lbrack h^{\alpha}_{\beta}, h^{\gamma}_{\zeta} \rbrack & = &
\delta_{\beta}^{\gamma} h^{\alpha}_{\zeta} -
\delta^{\alpha}_{\zeta} h_{\beta}^{\gamma} \, , \nonumber \\
\lbrack h^{\mu}_{\nu}, h^{\rho}_{\sigma} \rbrack & = & \delta_{\nu}^{\rho}
h^{\mu}_{\sigma} - \delta^{\mu}_{\sigma} h_{\nu}^{\rho} \, , \nonumber
\\ 
\lbrack h^{\alpha}_{\beta}, d_{\gamma \mu} \rbrack & = & 
-\delta^{\alpha}_{\gamma}
d_{\beta \mu} + \frac{1}{2} \delta^{\alpha}_{\beta} d_{\gamma
\mu} \, , \nonumber \\
\lbrack h^{\mu}_{\nu}, d_{\alpha \rho} \rbrack & = & -\delta^{\mu}_{\rho}
d_{\alpha \nu} + \frac{1}{2} \delta^{\mu}_{\nu}d_{\alpha \rho} \,
, \nonumber \\ 
\lbrack d_{\alpha \mu}, d_{\beta \mu} \rbrack & = &  \frac{1}{2} 
\varepsilon_{\alpha \beta} \varepsilon_{\rho \mu} h_{\nu}^{\rho}
 + \frac{1}{2} \varepsilon_{\mu \nu} \varepsilon_{\gamma
\alpha } h_{\beta}^{\gamma}.
\end{eqnarray}
and are normalized as
\begin{eqnarray}\label{norm}
\langle h^{\alpha}_{\beta}, h^{\gamma}_{\zeta} \rangle & = &
\delta_{\beta}^{\gamma} \delta^{\alpha}_{\zeta} - \frac{1}{2}
\delta^{\alpha}_{\beta} \delta^{\gamma}_{\zeta} \, , \nonumber \\ 
\langle h^{\mu}_{\nu}, h^{\rho}_{\sigma} \rangle & = & \delta_{\nu}^{\rho}
\delta^{\mu}_{\sigma} - \frac{1}{2} \delta^{\mu}_{\nu}
\delta^{\rho}_{\sigma} \, , \nonumber \\
\langle d_{\alpha \mu}, d_{\beta \nu} \rangle & = & - \frac{1}{2}
\varepsilon_{\alpha \beta} \varepsilon_{\mu\nu}.
\end{eqnarray}
The representation $ ad({\frak H})\frak M $ is obviously 
$ ( \underline{2} , \underline{2} ) $.   

Now we have to choose a gauge group and to define the homomorphism
$\lambda$ so that we get a nontrivial intertwining operator
$\Phi$, see equation (\ref{phi-map}).  We take ${\frak G} = A_{2m+1}$ 
and fix the
embedding of $su(2) \oplus su(2)$ into it by the decomposition of
the defining representation
\begin{equation}
 \underline{2m+2} \longrightarrow (\underline{2},\underline{1}) +
m \times (\underline{1}, \underline{2}).
\end{equation}
We take $ \{ f_{\alpha} , f_{\mu} \otimes g_{s} \} , \, \alpha =1,2 ,\, 
\mu =3,4 , \, s = 1 ,\ldots ,m $ as an adapted basis of 
$ R_{\underline{2m+2}} $, where $ su(2) \oplus su(2) $ acts 
trivially on $ \{ g_{s} \} $ and on 
$ \{ f_{\alpha} , f_{\mu} \} $ analogously as $ sp(2) \oplus sp(2) $ 
on $ \{ e_{\alpha} , e_{\mu} \} $ above.  
One can easily find that the centralizer of $su(2) \oplus su(2)$ in
$A_{2m+1}$ is $su(m) \oplus u(1)$, where $ H \in u(1) $ acts on $ \{
f_\alpha \} $ by multiplication with $ \frac{m}{m+1} $ and on $ \{ 
f_{\mu} \otimes g_{s} \} $ by multiplication with $ - \frac{1}{m+1} $.
The restriction of $ad
A_{2m+1}$ to $su(2) \oplus su(2) \oplus su(m) \oplus u(1)$ reads
\begin{eqnarray}\label{dec}
ad A_{2m+1} &=& (\underline{2m+2})^* \tilde \otimes (\underline{2m+2}) 
\longrightarrow \nonumber \\
&& (\underline{3},\underline{1},\underline{1})(0) +
(\underline{1},\underline{3}, \underline{1})(0) +
(\underline{1},\underline{1}, ad\, su(m) )(0) 
+ (\underline{1},\underline{1}, \underline{1})(0) +\nonumber\\
&& +  (\underline{1},\underline{3}, ad \,su(m) )(0) +
(\underline{2},\underline{2}, \underline{m})(-1)  +
(\underline{2},\underline{2}, \underline{m}^*)(1).
\end{eqnarray}
A basis of generators in $A_{2m+1}$ adapted to the decomposition
(\ref{dec}) can be taken to be $\{H^{\alpha}_{\beta},\,
H^{\mu}_{\nu}, \, H^i_k, \, H, \,  A^{\mu i}_{\nu k}, \,
D_{\alpha}^{ \mu i},\, D^{\alpha}_{ \mu i}\} \, , \,
{\alpha},{\beta} = 1,2,\, {\mu},{\nu}= 3,4,\, \, i,k = 1, \ldots,
m$, and all the generators have zero trace in any pair of indices
of the same kind.
The generators 
\begin{eqnarray}
 H^{\alpha}_{\beta} & = & f^\alpha \otimes f_\beta - \frac{1}{2} 
\delta^\alpha_\beta ( f^1 \otimes f_1 + f^2 \otimes f_2 )  \, \\ 
H^{\mu}_{\nu} & = & (f^\mu \otimes g^i ) \otimes (f_\nu \otimes g_i) 
-\frac{1}{2} \delta^\mu_\nu ( f^\rho \otimes g^i ) \otimes (
f_\rho \otimes g_i )   
\end{eqnarray} 
span two $su(2)$ subalgebras, whereas the generators 
\begin{equation} 
H^i_k = (f^\rho \otimes g^i ) \otimes ( f_\rho \otimes g_k ) 
- \frac{1}{m}\delta^i_k ( f^\rho \otimes g^l) \otimes (f_\rho
\otimes g_l ) 
\end{equation}  
span the $su(m)$; all of them  have the same commutation
relations as in (\ref{com}). 
Another relevant commutator is 
\begin{equation}
\label{commutator}
[D_{\alpha}^{ \mu i}, D^{\beta}_{\nu k}] =
\delta^{\beta}_{\alpha} A^{\mu i}_{\nu k} - \delta^{\mu}_{\nu}
\delta^i_k H^{\beta}_{\alpha}
+ \frac{1}{2}\delta^{\beta}_{\alpha} \delta^{\mu}_{\nu} H^i_k  
+  \frac{1}{m}\delta^{\beta}_{\alpha} \delta^i_k H^{\mu}_{\nu} 
- \frac{m+1}{2m}\delta^{\beta}_{\alpha} \delta^{\mu}_{\nu}
\delta^i_k H. 
\end{equation}

We emphasize that the generators $\{H^{\alpha}_{\beta},
\,H^{\mu}_{\nu}, \, H^i_k\}$ are normalized as follows
\begin{eqnarray}\label{norm1}
\langle H^{\alpha}_{\beta}, H^{\gamma}_{\zeta} \rangle & = &
\delta_{\beta}^{\gamma} \delta^{\alpha}_{\zeta} - \frac{1}{2}
\delta^{\alpha}_{\beta} \delta^{\gamma}_{\zeta} \, , \nonumber \\
\langle H^{\mu}_{\nu}, H^{\rho}_{\sigma} \rangle & = & m \bigl(
\delta_{\nu}^{\rho} 
\delta^{\mu}_{\sigma} - \frac{1}{2} \delta^{\mu}_{\nu}
\delta^{\rho}_{\sigma} \bigr) \, , \nonumber \\
\langle H^i_k, H^l_m \rangle & = & 2 \bigl( \delta^l_k\delta^i_m -
\frac{1}{m} \delta^i_k \delta^l_m \bigr) \, ,
\end{eqnarray}
which means that the second $su(2)$-subalgebra has index $m$, and
the $su(m)$-subalgebra has index $2$.

We see that the operator $\Phi$ intertwines
$ad({\frak H}){\frak M} = (\underline{2},\underline{2})$ with 
$(\underline{2},\underline{2}, \underline{m})(-1) $ and
$(\underline{2},\underline{2}, \underline{m}^*)(1)$ and is
parameterized by a complex vector $a^i,\, i = 1,2 \cdots , m$. 

Now we can define the homomorphism $\lambda$  putting 
\begin{equation}\label{hom}
\lambda'(h^{\alpha}_{\beta}) = H^{\alpha}_{\beta} \, , \quad 
\lambda'(h^{\mu}_{\nu}) = H^{\mu}_{\nu}.
\end{equation}
Then $\Phi$ takes the form  
\begin{equation}
\label{intertwiner}
 \Phi (d_{\alpha \mu}) = \frac{1}{2} a_i \varepsilon_{\nu \mu}
D_{\alpha}^{\nu i} + \frac{1}{2} a^i \varepsilon_{\beta \alpha}
D^{\beta}_{\mu i} \, , \quad  \bar a_i = a^i.
\end{equation}

Before we calculate the potential $ V $, see equation (\ref{orbitcurv}),
we make some remarks about the resulting gauge field. If $m \geq
1$ and $\mid a \mid ^2 = a_i a^i \neq 0$, the intertwining operator $\Phi$
maps $ \frak M $ into the $ su(4) $ subalgebra of ${\frak G} = su(2
m + 2) $, spanned by
\begin{eqnarray} 
\tilde H & = & (f^\rho \otimes g^i )
\otimes (f_\rho \otimes g_k) \frac{a_i a^k}{\mid a \mid ^2} -
f^{\gamma} \otimes f_{\gamma} \quad , \quad H^\alpha_\beta \, , \\
 \tilde H^\mu_\nu & = & \left((f^\mu \otimes g^i )
\otimes (f_\nu \otimes g_k) - \frac{1}{2} \delta^\mu_\nu ( f^\rho
\otimes g^i ) \otimes ( f_\rho \otimes g_k )\right) \frac{a_i a^k}{\mid a
\mid ^2} \, , \\
D_{\alpha}^{ \mu} & = & D_{\alpha}^{ \mu i} \frac{a_i }{\mid a\mid
} \, ,\quad D^{\beta}_{\nu} = D^{\beta}_{\nu k} \frac{ a^k}{\mid a
\mid } \, .  \\
\end{eqnarray} 
If $\mid a \mid ^2 = 1$, it extends the homomorphism $ \tau' $, defined by
\begin{equation}
{\tau}' ( h^\alpha_\beta ) = \lambda' ( h^\alpha_\beta ) = H^\alpha_\beta \, , 
\quad  {\tau}' ( h^\mu_\nu ) = \tilde H^\mu_\nu
\end{equation} 
to a homomorphism from $ sp(2) $ into $ su(4) $. 
Hence, the field strength, see equation (\ref{orbitcurv1}),  
takes values  only in an $ su(4) \oplus su(2) $
subalgebra of $ su(2 m + 2) $, where the $ su(2) $ subalgebra is given by 
$ {\tau_1}' ( {\frak H} ) $, the homomorphism $ {\tau_1}' $ being  
defined by 
\begin{eqnarray}
{\tau_1}' ( h^\mu_\nu ) &:=& H^\mu_\nu - \tilde H^\mu_\nu  \\ 
&=&  \left( (f^\mu  \otimes g^i ) \otimes ( f_\nu \otimes g_k )
- \frac{1}{2} \delta^\mu_\nu ( f^\rho \otimes g^i ) \otimes ( f_\rho \otimes g_k) 
\right)(\delta_i^k - a_i a^k) \, , \nonumber \\
{\tau_1}' ( h^\alpha_\beta ) &:=& 0.
\end{eqnarray}  

To calculate the potential $ V $, we have to fix the standard invariant
metric $\gamma_{K/H}$ on $S^4$. We choose this metric so that the
scalar product induced by it on ${\frak M}$ coincides up to the   
sign with  the restriction to ${\frak M}$ of the canonically
normalized invariant bilinear form $<, >$ on $sp(2)$.

Inserting the intertwining operator (\ref{intertwiner}) into
equation (\ref{orbitcurv}) and taking into
account that the elements $\{(d_{\alpha \mu} + d^{\alpha \mu}),\,
i(d_{\alpha \mu} - d^{\alpha \mu})\}, \, \alpha \leq \mu, \,
\alpha < 2$ form an orthonormal basis of ${\frak M}$, we get the
following potential
\begin{equation}\label{pot}
V(\varphi) = 12 (a_i a^i -1)^2 + 6(m-1) \, . 
\end{equation}
The first term of the potential comes from the $su(4)$ part of the gauge 
field and the second term comes from the $su(2)$ part.   

The potential has two extrema: the absolute minimum
(corresponding to the Higgs vacuum) with $a_i a^i =1$
and the local maximum, for which $a_i = 0$. 

First we will discuss the case $\mid a \mid ^2 = a_i a^i = 1 $. 
Having fixed $a_i$ with this property, we see that
the symmetry is broken to $ R_0 = SU(m-1) \times U(1)$. 
The first term of the potential $ V $  
vanishes, i.e. the $su(4)$ part of the gauge field 
is a pure gauge. Thus, we can choose a gauge so that this part 
of the gauge field vanishes and we arrive at the simple result that 
the gauge field on $ S^4 $ is $ su(2) \equiv {\tau_1}' ( {\frak H} )$-valued. 
Hence, this solution corresponds in the framework of symmetric 
gauge fields to the homomorphism $ {\tau_1}' : {\frak H} \rightarrow 
{\frak G} $ with vanishing intertwining operator $ \Phi $. 
The centralizer of $ {\tau_1}' ( {\frak H} ) $ in $ su(2 m +2) $ is 
$ {\frak C}_{{\tau_1}'({\frak H})} ( {\frak G} ) = su ( m-1 )
\oplus su(4) \oplus u(1) $ 
and we have $ {\frak R} = {\frak C}_{{\tau_1}'({\frak H})} ( {\frak G} ) $.

In the second case $ a_i = 0 \, , i= 1, \ldots , m $ 
we have an $ su(2) \oplus su(2) \equiv \lambda' ({\frak H} ) $-valued 
gauge field and $ {\frak R} = {\frak C}_{{\lambda}'({\frak H})} ( {\frak G} )
= su ( m ) \oplus u( 1) $.

Thus, we have found
the scalar field configurations, corresponding to the solutions of
the original Yang-Mills equations. The explicit form of these
solutions is given by formula (\ref{gauge field}), and it remains
to find $ \bar \theta _{ {\frak H }}$.

It is known that the group $K = Sp(2)$ can be represented by
 unitary quaternionic $2 \times 2$-matrices.
Then its Lie algebra $\frak K$ is parameterized as follows
\begin{equation}
{\frak K} = \left\{ \left(
\begin{array}{ll}
a & b \\ 
- \bar b & c
\end{array} \right) ;\, a , b , c \in {\Bbb H};\, \Re{a} = \Re{c} = 0 \right\} ,
\end{equation}
where $ {\Bbb H} $ is the skew field of quaternions, $ \bar a $ is the 
quaternionic conjugate and $ \Re{a} $ is the real part 
of the quaternion $ a $. 
The spaces $ \frak M $ and $ \frak H $ are given by
\begin{eqnarray}
{\frak M} &=& \left\{ \left(
\begin{array}{ll}
0 & b \\
-\bar b & 0 
\end{array} \right) ;\, b \in {\Bbb H} \right\} \quad , \\
{\frak H} &=& \left\{ \left(
\begin{array}{ll}
a & 0 \\
0 & c 
\end{array} \right) ;\, a , c \in {\Bbb H};\, \Re{a} = \Re{c} = 0 
\right\} \quad .
\end{eqnarray}
It is obvious that $ [{\frak H},{\frak M}] \subset {\frak M} $ and 
$ [{\frak M},{\frak M}] \subset {\frak H} $. 
We identify the upper $su(2)$-subalgebra with the one spanned by
$h^{\alpha}_{\beta}$ and the lower $su(2)$-subalgebra with the one
spanned by $h^{\mu}_{\nu}$.

Now we have to construct a section $ s_2 : K / H = S^4 
\rightarrow K = Sp(2) $ to calculate the form $ \bar \theta = s_2^* \theta $.
We can parameterize the group $ Sp(2) $ 
by $ u_1 , u_2 \in SU(2) $ and $ x \in {\Bbb H} $:
\begin{equation}
\frac{1}{\sqrt{1 + x \bar x}}  
\left( \begin{array}{ll}
1 & x \\
- \bar x & 1
\end{array} \right) 
\left( \begin{array}{ll}
u_1 & 0 \\
0 & 1
\end{array} \right)
\left( \begin{array}{ll}
1 & 0 \\
0 & u_2
\end{array} \right) \in {Sp(2)} \quad .
\end{equation}
This parameterization is not valid for infinite $ x $, but that is clear 
because $ K / H = S^4 $. We identify $ {\Bbb H} \cup {\infty} $ with $ S^4 $ 
and choose the section $ s_2 $ by putting 
\begin{equation}
s_2 ( x ): = \frac{1}{\sqrt{1 + x \bar x}}
\left( \begin{array}{ll}
1& x \\
- \bar x & 1
\end{array} \right) \quad .
\end{equation}
After a short calculation we get 
\begin{eqnarray}
\bar \theta &=& s_2^* \theta = (s_2(x))^{-1} d (s_2(x)) \nonumber \\
&=& \frac{1}{2 (1+x \bar x)} \left( \begin{array}{ll}
x d \bar x - d x \bar x & dx \\
- d \bar x & \bar x d x - d \bar x x
\end{array} \right) \quad , \\
\bar \theta_{{\frak H}} &=& \frac{1}{2 (1+x \bar x)} \left( \begin{array}{ll}
x d \bar x - d x \bar x & 0 \\
0 & \bar x d x - d \bar x x
\end{array} \right) \, .
\end{eqnarray}
Taking into account the relation between quaternions and Pauli matrices, 
we see that the two $ su(2) $-components of $\bar \theta_{{\frak H}}$ 
are the fundamental instanton resp.  anti instanton connections on 
$ S^4 $. The gauge field $A$, see equation (\ref{gauge field}), 
reads
\begin{eqnarray}
A &=& 
- \frac{\eta(t) \quad {\cal E}}{(n-1) r^{(n-1)}} dt 
+ \psi' ( \bar \theta_{{\frak H}} ) \, , 
\end{eqnarray}
where $ {\cal E} \in {\frak R} $ and $ \psi' = {\tau_1}' $ in the 
first and $ \psi' = {\lambda}' $ in the second case and we see that  
the $ S^4 $ part of the gauge field is an instanton like solution. 
In particular, if we put $m =0$, the intertwining operator $\Phi$
is identically zero, and the centralizer of $ \lambda'
(\frak H ) $ in 
$ \frak G $ is  trivial in this case, so that $ \hat A $ has to
be zero, and the gauge field on 
$M \times S^4 $ consists only of the instanton gauge field on
$S^4$. 

Hence, our approach can also lead to noncommutative solutions 
of EYM-systems,  if the stabilizer group is  not $ U(1) $.

One more remark is in order. It is easy to see that our method of
finding solutions of EYM-equations makes it possible to discuss 
stability: It is clear that solutions corresponding   
to the local maximum of the Higgs potential are unstable.
This means that  the full solutions with dynamical gravitational 
field are also unstable, because the YM degrees of freedom can not 
be compensated by the gravitational degrees of freedom. 
On the other hand solutions corresponding to the Higgs vacuum 
can be considered as stable, if we treat gravitation  
as a fixed background field. But, of course, to discuss stability 
in the full dynamical context is very complicated. One can use 
for instance the concept of linear stability as discussed in 
\cite{Straumann494}.

\section*{Acknowledgments}

One of the authors (I.V.) is grateful to the Center of Natural Sciences of
the University of Leipzig for the warm hospitality extended to him during
his stay in Leipzig. He also acknowledges partial support under the
INTAS-93-1630 project.

\end{document}